\documentstyle[12pt,fleqn]{article}
\newcommand{\preprint}[1]{\hfill{\sl preprint - #1}\par\bigskip\par\rm}
\def\titolo{\par\bigskip\begin{center}\bf\LARGE}
\def\endtitolo{\end{center}\par\bigskip\par\rm\normalsize}
\def\instit{\begin{center}\large}
\def\endinstit{\end{center}\rm\normalsize}
\def\references{\begin{thebibliography}{10}}
\def\endreferences{\end{thebibliography}\end{document}}
\newcommand{\dip}{\smallskip Dipartimento di Fisica,
                                Universit\`a di Trento}
\newcommand{\infn}{\smallskip Istituto Nazionale di Fisica Nucleare,\\
                                 Gruppo Collegato di Trento,\\ 38050 Povo (TN)
Italia}
\newcommand{\dinfn}{\dip\\ and \infn}
\newcommand{\btit}{\begin{titolo}}
\newcommand{\etit}{\end{titolo}}

\newcommand{\Idinfn}{\begin{instit}\dinfn\end{instit}}

\renewcommand{\author}[1]{\begin{center}\Large #1\end{center}}
\renewcommand{\date}[1]{\par\bigskip\par\sl\hfill #1\par\medskip\par\rm}
\newcommand{\email}[1]{e-mail: \sl #1@science.unitn.it,
                               #1@itncisca.bitnet, 37953::#1\rm}
\newcommand{\femail}[1]{\footnote{\email{#1}}}
\newcommand{\pacs}[1]{\smallskip\noindent{\sl PACS number(s):
                       \hspace{0.3cm}#1}\par\bigskip\rm}
\newcommand{\babs}{\hrule\par\begin{description}\item{Abstract: }\it}
\newcommand{\eabs}{\par\end{description}\hrule\par\medskip\rm}
\newcommand{\ack}[1]{\par\section*{Acknowledgements} #1}
\renewcommand{\vec}[1]{{\bf #1}}       %%%  vectors in bold
\newcommand{\nn}{\nonumber}            %%%  no number for eqnarray
\newcommand{\beq}{\begin{eqnarray}}    %%%  begequation/eqnarray
\newcommand{\eeq}{\end{eqnarray}}      %%%  endequation/eqnarray
\newcommand{\beqn}{\begin{eqnarray}}   %%%  begequation/eqnarray
\newcommand{\eeqn}{\end{eqnarray}}     %%%  endequation/eqnarray
\newcommand{\at}{\left(}               %%%  open (
\newcommand{\ct}{\right)}              %%%  close )
\newcommand{\be}{\beta}

\newcommand{\ep}{\varepsilon}

\begin{document}

%%%% \tableofcontents       %%%%%%   index of section

\preprint{UTF 363   \\ gr-qc/9510016}

\btit
%%%%%%%%%%%%%%% inserire qui il titolo %%%%%%%%%%%%%%%%%%%%
 Hessling's Quantum Equivalence Principle and the
Temperature of an Extremal Reissner-Nordstr\"{o}m Black Hole
\etit

\author{Valter Moretti \femail{moretti}}

\Idinfn

\date{October 1995}

\babs \\
The Hessling improvement of the Haag, Narnhofer and Stein
principle is analysed in the case of a massless scalar field
propagating outside of an extremal R-N black
hole. It is found that this sort of
 ``Quantum (Einstein's) Equivalence Principle''
selects only the
R-N vacuum  as a physically sensible state, i.e.,
it selects the  temperature $T=0$ only.
\eabs

%%%%%%%%%%%%%%%%  Corpo dell' articolo %%%%%%%%%%%%%%%%%%%%
In this note we would like to
 report some  results obtained employing the
Hessling principle (see below) in order to select the
physically sensible KMS states of a massless
scalar field propagating outside of
 an extremal R-N black hole. Explicit calculations and more extended
discussions  will appear in the changed and final version of ref.
 \cite{moretti}.\\

To begin with, we recall that
 Haag, Narnhofer and Stein in \cite{haag}
 proved that if one assumes fairly standard axioms of
 quantum (quasi-free) field theory, particularly {\em local definiteness} and
 {\em local stability} in an at least stationary,
 causally complete space-time region,
 the thermal Wightman functions
  in the  {\em interior} of this region will transform  into {\em non thermal}
 and {\em massless} Wightman functions in the Minkowski space-time
 when the geodesic
 distance of the arguments is  vanishing.\\
We  called \cite{moretti} this statement, which from a naive
point of view seems
 to follow from the Einstein equivalence principle, the HNS {\em theorem}.
It is important to stress that
 the   statement above is valid also in the case $T=0$. \\
For example, in case of two-point thermal Wightman functions of a
 scalar field  it holds:
 \beq
  {\lambda}^2 W^{\pm}_{\be}(x+\lambda z_{1},x+\lambda z_{2}) \rightarrow
\frac{1}{4\pi^{2}}
\frac{1}{g_{\mu \nu}(x)z^{\mu}z^{\nu}}
   \;\;\; \mbox{as} \;\;\; \lambda \rightarrow 0^{+}
 \:, \label{HNS interno}
\eeq
where $z=z_{2}-z_{1}$.
In the equation above the coordinates $x \equiv x^{\mu} $
indicate a point in the {\em interior}
of the region,
\beq
z_{(j)} \equiv z_{(j)}^{\mu}
\frac{\partial \;\;}{\partial x^{\mu}}\mid_{x} \nn
\eeq
 indicates  vectors
 in the {\em tangent space} at $x$
  and finally  we used the obvious
 notation:
\beq
x+z_{(j)} \equiv x^{\mu}+z_{(j)}^{\mu}\:.\nonumber
\eeq
We stress  that both sides of Eq.(1) are
{\em distributions} acting on
a couple of smooth test functions in the corresponding variables
$z_{1}$ and $z_{2}$.\\
 Conversely, the  Haag, Narnhofer and Stein {\em principle} \cite{haag}
generalises
 the HNS theorem and it affirms
 that in the case the  space-time region we are dealing with
 is
 just {\em a part}
 of the whole
 manifold separated by {\em event horizons},
 the  point coincidence behaviour
 of the  Wightman functions for a physically
 sensible (~thermal {\em or not} ) state must hold also
{\em onto the  horizons}.\\
 Haag, Narnhofer and Stein proved
 in \cite{haag} that in the case of Rindler and Schwarzschild space-times,
 this constraint holds {\em  only} for $\be_{T}= 1/T_{U,H}$, where $T_{U}$
and $T_{H}$ are
the {\em Unruh} and {\em Hawking temperatures}.\\
 Eq.~(\ref{HNS interno})
 implies:
 \beq
  {\lambda}^2 W^{\pm}_{\be}(x,x+\lambda z) \rightarrow \frac{1}{4\pi^{2}}
\frac{1}{g_{\mu \nu}(x)z^{\mu}z^{\nu}}
   \;\;\; \mbox{as} \;\;\; \lambda \rightarrow 0^{+} \:, \label{HNS interno2}
\eeq
where $x$ is a point on the horizon.
   Following the original paper of Haag Narnhofer and Stein, we interpreted
 Eq.~(\ref{HNS interno}) in
 \cite{moretti}
 in this weaker sense  and thus we
 evaluated the limit in Eq.~(\ref{HNS interno2})
 for {\em real} vectors $z$, space-like or
 time-like; for light-like vectors we expected a divergent
 limit.\\
 In order to  use the (weak) HNS principle for two-point Wightman functions
 one has
 to check  their behaviour as one point is fixed on the horizon and the
 other is running toward the first from the inside of the considered region.
 This is not as simple as one might think at first, because
 the  metric could become degenerated  on the horizon in the stationary
 coordinates which
 define the studied thermal Wightman functions, consequently it
 could not be possible to write down the right hand side of Eq.~(2) in that
 coordinate frame.
 However, as pointed by Haag, Narnhofer and Stein in \cite{haag},
 one can check the
 validity of the HNS principle in stationary coordinates
 using directly Eq.~(2)  for  specific points on the  horizons, i.e.,
 for those which
 belong to the intersection of the past and the future horizons, and along
 appropriate directions (along the space-like geodesics),
 but not on the whole horizon. The check
 of the behaviour of Wightman functions in  these
 ``few'' points  is sufficient to determine the Unruh and the
 Hawking temperatures respectively in the case of the Rindler wedge
 and the Schwarzschild background.
 Hessling proved in \cite{hessling} that the HNS principle, in the case of
 the Rindler wedge, determines
 the Unruh temperature only when it works on the intersection of the horizons,
 but it does not determine any temperature by
 considering the remaining points. In \cite{moretti} we  reported an
 independent proof of this fact.\\
 Unfortunately in  the case of an extremal Reissner-Nordstr\"{o}m black hole
 the past and the future horizons {\em do not intersect} (see \cite{moretti}
 and references therein) and furthermore
 it is not possible to deal with the original procedure of Haag, Narnhofer
 and Stein because in this case some important technical
 hypothesis does not hold,
 e.g., the requirement  of a
{\em  non-vanishing surface gravity} or related parameters
 \cite{haag}.\\
In such a situation the Hessling development of the HNS principle
 results to be
 very useful because, in the Minkowski space-time at least,
 it determines the Unruh temperature  working also
 considering  the points which do not belong to the intersection of horizons
 \cite{hessling}.  In \cite{moretti} we   reported an independent
 proof of this fact, too.
 We  expected a similar result in the case of an extremal  R-N black hole
 where the intersection does not exist.\\
 Furthermore \cite{hessling} one could note that, in a
 real (Schwarzschild)  black hole, the {\em past} event
 horizon does not exist and thus also the {\em intersection}
 of horizons does not exist. For this reason the Hessling principle
 result to be
 very important as far as the possibility to use this in more physical
 situations than the {\em eternal} black hole cases is concerned.\\
 {\em Hessling's principle}, trying to define a
{\em  Quantum (Einstein) Equivalence Principle} to be imposed on the
 physically sensible quantum states
\cite{hessling}\footnote{Really, Hessling considers
 in \cite{hessling} also $n$-{\em point functions}, but we  restricted
\cite{moretti}
 our discussion by considering only the case of a scalar {\em quasifree}
 field and thus by studying only the {\em two-point} functions.},
 requires the {\em existence} of the limit:
 \beq
 \lim_{\lambda\rightarrow 0^{+}}
 N(\lambda)^{2}\: W^{\pm}_{\be}(x+\lambda z_{1},x+\lambda z_{2}) \:,
 \label{hessling1}
 \eeq
as a {\em continuous} function of $x$,
 for some function $N(\lambda)$ monotonous and nonnegative for $\lambda>0$.
 Furthermore, it requires the validity of a much more strong condition
 in every {\em local inertial coordinate system around}
 $x$ (i.e., a coordinate frame such that
 $g_{\mu\nu}(x)=diag(-1,1,1,1)$, $\partial_{\rho} g_{\mu\nu}(x)=0$):
 \beq
 \lim_{\lambda\rightarrow 0^{+}}\frac{d\:\:}{d\lambda}
 \: N(\lambda)^{2} W^{\pm}_{\be}(x+\lambda z_{1},x+\lambda z_{2}) = 0\:.
 \label{hessling2}
 \eeq
 The latter requirement
 is connected with the fact that within a local inertial
 system the metric looks like the Minkowski metric {\em up to the first order}
 in the coordinate derivatives. In the Minkowski
 background and using Minkowskian coordinates
 Eq.(\ref{hessling2}), which involves coordinate derivatives up to
{\em the first order}, results to be satisfyed and thus we expect this
will hold in curved backgrounds by using local inertial coordinate
systems\footnote{We stress that Hessling generalises the above
  equations in order to be able to use  a {\em non} local inertial coordinate
  system \cite{hessling}, too.},
too (see \cite{hessling} for details).
We can note that, away from the
horizons, the validity HNS {\em theorem} implies the validity
of the {\em first}
 Hessling requirement with $N(\lambda)=\lambda $.
Furthermore, the validity of the HNS {\em principle on a horizon} implies
the validity of the {\em first} Hessling requirement there.\\
As in the case of the HNS principle, in \cite{moretti},
we used the Hessling principle
 on an event horizon in a {\em weaker}
 version,
by checking the validity
of Eq.(2) as well as of the following equation:
\beq
\frac{d\:\:}{d\lambda}
{\lambda}^2 W^{\pm}_{\be}(x,x+\lambda z) \rightarrow 0
   \;\;\; \mbox{as} \;\;\; \lambda \rightarrow 0^{+} \:, \label{hessling}
\eeq
where $x$ belongs to the horizon, $z$ is a {\em real space-like or time-like}
 vector  and a  {\em local inertial
coordinate system} is used.\\
It is interesting to check whether
 the HNS principle and the Hessling principle select special
 temperatures for thermal states in the  case of
 an extremal Reissner-Nordstr\"{o}m black hole (see \cite{moretti} for
 a more extended discussion involving other methods, too).
It is possible to check the HNS and the Hessling principles by studying
the behaviour of two point Wightman functions along every geodesic which starts
from the event horizons.  We  note that for every geodesic which meets
the horizon in a point $x$ there is a locally geodesic coordinate frame
with
the origin at  $x$. Furthermore we can
choose the  geodesic to be a  coordinate
axis, the running coordinate being just the length of the  geodesic measured
from $x$.
By varying the  geodesics which meet the horizon one obtains all the possible
points of the horizon together with  their tangent vectors.
 We stress that it is possible to execute this checking procedure also
by using  stationary
coordinates, because the geodesic length is invariant
and so is not sensitive to the degeneracy of this {\em representation} of the
metric. \\
Using the above method in the case of the {\em Rindler
wedge} we concluded in \cite{moretti} that
 only the points in the {\em intersection} of horizons
 really select a  temperature of thermal states by the {\em HNS principle}:
 the {\em Unruh temperature}
$T_{U}=1/\beta = 2\pi$.
On the other hand,
 considering the points on the horizon
which
do not belong to the intersection of horizons,
we obtained in \cite{moretti}
 that the {\em Hessling principle} holds {\em only if} the
temperature is the
{\em Unruh temperature} once again\footnote{We also   observed
 that also the limit
 case $\beta \rightarrow + \infty$ {\em does not satisfy}
 the Hessling condition.}.
The same results appear on \cite{hessling}.\\
In case of a massless
 scalar field in the 4-dimensional Reissner-Nordstr\"om background,
the metric we are interested in reads:
\beq
ds^2=-\at1-\frac{R_H}{R}\ct^2\,(dx^0)^2+
\at1-\frac{R_H}{R}\ct^{-2}\,dR^2+R^2\,d\Omega_{2}
\:,\label{bh}
\eeq
where we used polar coordinates, $R$ being the radial one and
$d\Omega_{2}$ being  the metric of the unit-$2$-sphere.
The horizon radius is $R_H=MG=Q$,
$M$ being the mass of the black hole, $G$ the Newton constant and $Q$
its charge.
It is obvious that any redefinition of {\em space-like coordinates} does not
change the thermal properties of our field theory because these properties
depend on the  time-like coordinate and in particular on its
tangent Killing vector. Thus, as in \cite{cvz},
 we  redefined the Reissner-Nordstr\"{o}m radial coordinate
in order to obtain the following approximated form of the metric
{\em near the horizon}, i.e., $r\rightarrow +\infty $:
\beq
d s^2=\frac{1}{r^{2}}\left[ -dt^2
+dr^2+r^{2}d\Omega_{2}\right]  \label{metrica}\:,
\eeq
or
\beq
d s^2=\frac{1}{\vec{x}^{2}}\left[ -dt^2
+(d\vec{x})^{2}\right]  \label{metrica2}\:,
\eeq
where we used the obvious notation
\beq
t=x^{0} \:,\nonumber
\eeq
\beq
\vec{x}\equiv (x^{1},x^{2},x^{3})\:\:\:,\:\:\: r=|\vec{x}| \:.\nonumber
\eeq
The previous metric is called the  {\em Bertotti-Robinson metric}
\cite{bertotti robinson}.
Note that this metric is {\em conformal}
 to Minkowski metric by the factor
$1/r^{2}$, {\em singular} at the origin; however we were interested  in the
region near
the horizon, i.e. $r \rightarrow +\infty$, and  this singularity is
absent there.\\
In order to perform explicit computations, we  considered
the large mass limit  of the black hole, i.e., $R_{H}\rightarrow +\infty$.
 In this limit, the whole region outside
of the black hole ($R > R_{H}$)
 tends to approach the horizon and thus
the above approximated metric tends to hold everywhere.\\
In \cite{moretti}, we calculated the geodesics which reach the past ($H^{-}$)
or the future ($H^{-}$) horizon of the considered metric.
We obtained {\em three} families of  geodesics which start from the
horizons.\\
Really,  a further family of time constant geodesics which {\em seem}
to reach the intersection of horizons $H^{+}\cap H^{-}$ appeared, too.
We stressed in \cite{moretti} that these geodesics {\em just seem}
 to reach the horizons
as $s \rightarrow \pm
\infty$, where $s$ is a Riemannian length calculated from
a point away from the horizon,
 but that is an obvious contradiction; actually, as discussed in
\cite{moretti},
 this is a consequence of the fact  that $H^{+}\cap H^{-}=\emptyset$.\\
We report on the {\em first} family only here, a complete description
appear in \cite{moretti}.\\

{\em First family of  geodesics which start from the horizons.}
\begin{eqnarray}
r&=&\frac{A}{ s}\:,\label{bunodegenere}\\
t&=&t_{0}\pm \frac{A}{s} \:,\label{bduedegenere}\\
\phi&=& s + \phi_{0}\label{btredegenere} \:,
\end{eqnarray}
where $A>0$, $t_{0}$, $\phi_{0}$ are real numbers.
The origin of the parameter $s$ is
chosen in a way that  the starting
point of the geodesic is on the  horizon.
This horizon will be  $H^{+}$ if the sign in front of $A$
in Eq.~(\ref{bduedegenere})
is $+$, otherwise it will be  $H^{-}$ if the sign is $-$.\\
All these geodesics are {\em space-like}.\\

Furthermore,
we calculated the Wightman function for a massless scalar field in the
metric
(\ref{metrica}) and checked the HNS principle as well as the Hessling principle
on the horizons.
This metric is {\em conformal} to the Minkowski metric, thus
 we were able to use the Dowker and Schofield's
method \cite{dowker1,dowker2} which connects the Wightman  functions
(in general the Green functions) of a
 scalar field in a static
manifold with the corresponding Wightman functions of the field in another,
{\em conformally}
related static manifold.\\
We found \cite{moretti} the thermal
Wightman functions of a massless scalar field
(minimally as well as  conformally coupled)
 propagating outside of a large mass, extremal
Reissner-Nordstr\"{o}m black-hole:
\beq
W^{\pm}_{\be}=   \frac{\mid \vec{x}\mid  \mid
\vec{x^{'}}\mid}{4\pi^2}\frac{\pi\
 \left\{\coth\frac{\pi}{\be}
(|\vec{x}-\vec{x}^{'}|+t-t^{'}\mp i\ep)+\coth
\frac{\pi}{\be}(|\vec{x}-\vec{x}^{'}|-t+t^{'}\pm i\ep)\right\} }{2\be
|\vec{x}-\vec{x}^{'}|} \label{wightman}
\eeq
By taking the limit $\mid\vec{x}\mid\rightarrow +\infty$  we
obtained the thermal Wightman functions calculated on the horizon
(in the argument $x$). In order to  to calculate this limit we had to
increase (to reach $H^{+}$) or decreased (to reach $H^{-}$)
the variable $t$ together the variable $\vec{x}$ and $\phi$ along the
geodesics obtained above.
 We  dealt with
$H^{+}$ only, because of the time symmetry of the problem.\\
 By considering either
the space-like
or the time-like geodesics which reach the future horizons
 we   produced the same function:
\beq
W^{\pm}_{\be}\left(x_{H^{+}},x^{'} \right)=
\frac{r^{'}\pi \left\{ 1 + \coth\left[\frac{\pi}{\be}\left( t^{'}-t_{0}-r^{'}
\cos( \phi^{'}-\phi_{0})
\right) \right]   \right\} }{4\pi^{2} 2 \be}
\label{ddieci}
\eeq
In order to check the HNS principle on $H^{+}$  for a space-like
 vector $z$ tangent to a  geodesic of the {\em first} family in the horizon,
we considered Eq.~(\ref{ddieci}) and
substituted  the variables $r^{'}$, $t^{'}$, $\phi^{'}$
for the   functions defined in the right hand side
of equations (\ref{bunodegenere}), (\ref{bduedegenere}), (\ref{btredegenere}),
using also the identifications: $t_{0}=t_{0}^{'} $ and $\phi_{0}=\phi_{0}^{'}$.
 Finally we redefined $s=\lambda z^{i}$.\\
We obtained, as $\lambda \rightarrow 0$:
\beq
\lambda^{2}W_{\be}\left(x_{H^{+}},x_{H^{+}}+\lambda z \right)\sim
\nonumber
\eeq
\begin{eqnarray}
&\sim&
\frac{A\pi \lambda^{2}}{4\pi^{2}2\be \lambda z^{i}}
\left\{ \coth\left[\frac{\pi}{\be}\left( A\frac{1}{ \lambda z^{i}}-
\frac{A \cos\left( \lambda z^{i}\right)}{\lambda z^{i}}\right)
\right] +1  \right\} \nonumber \\
&\sim& \frac{\lambda^{2}}{4\pi^{2}2 z^{i}}\frac{z^{i}}{\frac{1}{2}
(\lambda z^{i})^{2}}=\frac{1}{4\pi^{2}} \frac{1}{g_{\mu\nu}
\left( H^{+}\right) z^{\mu}z^{\nu}}\:. \nonumber
\end{eqnarray}
We concluded that in that case the HNS principle holds
 for any value of
$\be$. However the remaining two
 families of geodesics
produced the same result (see \cite{moretti} for explicit
calculations).\\
In \cite{moretti}, we were also able to  prove that the same result comes out
in the limit case of the R-N vacuum, i.e., $T=0$.\\
Finally, we  checked the {\em Hessling principle} in the final
version of \cite{moretti}.\\
Considering the space-like geodesics of the {\em first family} in Eq.s
(\ref{bunodegenere}), (\ref{bduedegenere}) and  (\ref{btredegenere}),
it arose by using also Eq.(\ref{ddieci}):
\beq
\frac{d\:\:}{d\lambda} \:
 \lambda^{2} W_{\be}(x_{H^{+}},x_{H^{+}}
 +\lambda z)  =  \frac{\Gamma}{\beta}\:
\frac{d\:\:}{d\lambda} \:
 \lambda
\left\{ 1+ \coth \left[\frac{\pi A}{\beta \lambda z} (1- \cos \lambda z)
\right] \right\} \:, \nn
\eeq
where the factor $\Gamma:= A/8\pi z$ is {\em finite}
 and it does not depend on $\lambda$ and $\beta$.\\
Some trivial calculations led to:
\beq
\frac{d\:\:}{d\lambda} \:
 \lambda^{2} W_{\be}(x_{H^{+}},x_{H^{+}}
 +\lambda z)  =  \frac{\Gamma}{\beta}
\left\{ 1+ \frac{X + \sinh X\: \cosh X -
\frac{\pi A}{\beta z} \sin \lambda z}{\sinh^{2}X}  \right\}\:, \nn
\eeq
where we also posed:
\beq
X := \frac{\pi A}{\beta z} \frac{1- \cos \lambda z}{\lambda}\:. \nn
\eeq
 Doing the limit as $\lambda \rightarrow 0^{+}$:
\beq
\frac{d\:\:}{d\lambda} \:
 \lambda^{2} W_{\be}(x_{H^{+}},x_{H^{+}}
 +\lambda z) \rightarrow \frac{\Gamma}{\beta}\:.\nn
\eeq
{\em This fact was sufficient to prove that the Hessling principle excludes
every
finite value of $\beta$}.\\
The limit case $T=1/\beta= 0$ survived only. Using the remaining two families
of geodesics, the
calculations resulted to be very similar and the same limit value
of the temperature survived. Furthermore,
 we   proved that the Wightman functions
of the R-N vacuum, i.e., the limit case $T=1/\be=0$, satisfies the Hessling
principle by considering directly the expression of the Wightman functions
of the R-N vacuum \cite{moretti}.\\

The most  important  conclusion which follows from the
calculations in \cite{moretti} is that {\em the (weak) HNS principle,
 in the case of an extreme Reissner-Nordstr\"{o}m
black hole, holds for every value of $\be$,
 i.e., it agrees with
the  method based on the
 elimination of the singularities of the  Euclidean manifold}
(see \cite{moretti} and ref.s therein), {\em  but
 the Hessling principle selects only the  null
 temperature, i.e., the R-N vacuum as a physical state.}\\
However, we report  some other important remarks.\\
Another important point is that we dealt with
  the limit of a large mass black
 hole and with a massless field, but we think that our conclusions
 should hold without to assume these strong conditions, too
(see the discussion on this in  \cite{moretti}).\\
Finally, we stress that recently P.H. Anderson, W.A. Hiscock and D.J
Loranz \cite{AHL}, by using of the metric (7) and the
Brown-Cassidy-Bunch formula (see \cite{AHL,rindler} and references therein)
argued
(and numerically checked by using the {\em complete}
 R-N metric) that the
Reissner-Nordstr\"{o}m vacuum state is the {\em only} thermal
state with a non-singular {\em renormalized} stress-tensor on the horizon of an
extremal R-N black hole.\\
We observe that  the ``improved'' {\em HNS prescription}, i.e. the
{\em Hessling principle}  agrees completely with the result of
Anderson Hiscock and Loranz, in particular it selects a state carrying
a renormalized stress-tensor {\em finite} on the horizon.
This fact comes out also both in the Rindler space where the HNS and
Hessling's
prescriptions select the Minkowski vacuum which has a regular stress tensor
on the horizon or in the Schwarzschild space where the HNS principle
selects the Hartle-Hawking state with the same property on the horizon
\cite{rindler}.

%%%%%%%%%%%%%%%%%%%%%%%%%%%%%%%%%%%%%%%%%%%%%%%%%%%%%%%%%%%%%%%%%%%%%%%%%%%%%%

%\section{       }

%%%%%%%%%%%%%%%%%%%%%%%%%%%%%%%%%%%%%%%%%%%%%%%%%%%%%%%%%%%%%%%%%%%%%%%%%%%%%%%

%\section{       }

%%%%%%%%%%%%%%%%%%%%%%%%%%%%%%%%%%%%%%%%%%%%%%%%%%%%%%%%%%%%%%%%%%%%%%%%%%%%%%%

%\section{Discussion}

%%%%%%%%%%%%%%%%%%%%%%%%%%%%%%%%%%%%%%%%%%%%%%%%%%%%%%%%%%%%%%%%%%%%%%%%%%%%%

\ack{I would like to thank G.Cognola, M.Toller, L.Vanzo
 and S.Zerbini of the Dipartimento di Fisica dell'Universit\`{a} di
 Trento for all the useful discussions about the topics of this  paper.\\
 I am grateful to Bernard Kay and a  referee of
 {\em Classical and Quantum Gravity}
 for the article in [1] who pointed out the Hessling paper to me.\\}

\end{document}